\documentclass[aps,showpacs,prl,
reprint,
  superscriptaddress]{revtex4-1}
\usepackage{graphicx}
\usepackage{natbib}
\usepackage{amsmath}
\usepackage{epstopdf}
\usepackage{color}
\usepackage{upgreek}
\usepackage{gensymb}

\definecolor{dgreen}{rgb}{0.0, 0.5, 0.0}

\usepackage{dsfont}
\usepackage{xspace}
\usepackage{color}
\usepackage[usenames,dvipsnames]{xcolor}
\usepackage[colorlinks=true,linkcolor=blue,urlcolor=blue,citecolor=blue]{hyperref}
\usepackage[normalem]{ulem}

\usepackage{mathrsfs}
\usepackage{fullpage}
\usepackage{amssymb}
\usepackage{graphicx}
\usepackage{amsmath}
\usepackage[strict]{changepage}
\usepackage{framed}
\usepackage{amsthm}
\usepackage{bbm}
\usepackage{colortbl}
\newcommand{\etal}{{\it et al.}}

\begin{document}
\newcommand{\unit}[1]{\,\mathrm{#1}}
\newcommand{\Figureref}[1]{Fig.~\ref{#1}}
\newcommand{\Eqnref}[1]{equation~(\ref{#1})}
\newcommand{\affilOU}{Homer L. Dodge Department of Physics and Astronomy, The University of Oklahoma,
Norman, OK 73019, USA}
\newcommand{\affilITAMP}{ITAMP, Harvard-Smithsonian Center for Astrophysics, Cambridge, MA 02138, USA}
\newcommand{\affilSandia}{Sandia National Laboratories, Albuquerque, NM 87185, USA}
\newcommand{\affilUNLV}{Department of Physics and Astronomy, University of Nevada Las Vegas, Las Vegas, NV 89154, USA}
\newcommand{\affilWWU}{Department of Physics and Astronomy, Western Washington University, Bellingham, WA 98225, USA}
\newcommand{\affilNavy}{Department of Physics, The United States Naval Academy, Annapolis, MD 21402, USA}
\title{Electric field cancellation on quartz: a Rb adsorbate induced negative electron affinity surface}
\author{J. A. Sedlacek}
\affiliation{\affilOU}
\author{E. Kim}
\affiliation{\affilUNLV}
\author{S. T. Rittenhouse}
\affiliation{\affilWWU}
\affiliation{\affilNavy}
\author{P. F. Weck}
\affiliation{\affilSandia}
\author{H. R. Sadeghpour}
\affiliation{\affilITAMP}
\author{J. P. Shaffer}
\email[]{shaffer@nhn.ou.edu}
\affiliation{\affilOU}
\date{\today}

\begin{abstract}
We investigate the (0001) surface of single crystal quartz with a submonolayer of Rb adsorbates.  Using Rydberg atom electromagnetically induced transparency, we investigate the electric fields resulting from Rb adsorbed on the quartz surface, and measure the activation energy of the Rb adsorbates.  We show that the adsorbed Rb induces a negative electron affinity (NEA) on the quartz surface.  The NEA surface allows low energy electrons to bind to the surface and cancel the electric field from the Rb adsorbates.  Our results are important for integrating Rydberg atoms into hybrid quantum systems and the fundamental study of atom-surface interactions, as well as applications for electrons bound to a 2D surface.
\end{abstract}

\maketitle
Due to recent technological advances in fabrication and trapping, hybrid quantum systems (HQS) consisting of atoms and surfaces, as well as electrons and surfaces, are fast emerging as ideal platforms for a diverse range of studies in quantum control, quantum simulation and computing, strongly correlated systems and microscopic probes of surfaces.  Miniaturization of chip surfaces is necessary to achieve large platform scalability, but decoherence and noise emerge as serious challenges as feature sizes shrink \cite{Behunin2012,Blatt2015,carter2013}.  Mitigating the noise is a fundamental and necessary step in realizing the full potential of HQSs for quantum technologies.

Combining ultracold Rydberg atoms with surfaces for HQS is attractive because Rydberg atoms can have large sizes, significant electric dipole moments and strong interactions. There have recently been a host of theoretical proposals for utilizing Rydberg atoms near surfaces  \cite{petrosyan2009reversible, harald2013, pritchard2014hybrid, kurizki2015quantum, hinds1997atoms}. Progress on the experimental front has been hampered by uncertainties in characterizing interactions of atoms with surfaces, although some recent work in this regards are noteworthy  \cite{Haroche2014, thiele2014manipulating, teixeira2015microwave}.

To take full advantage of Rydberg atom HQSs, a more complete understanding of surfaces is needed.  One problem is that Rydberg atoms incident upon metal surfaces can be ionized \cite{dunningSurface2000, softley2011}.  A second major hurdle is the background electric fields (E-fields) caused by adsorbates \cite{kubler2010coherent, Martin2012, hattermann2012detrimental, chan2014adsorbate, Spreeuw2010, Abel2011}.  Rydberg states are sensitive to adsorbate E-fields because they are highly polarizable \cite{gallagherBook}.  Adsorbate E-fields have caused problems for other experiments as well, including Casimir-Polder measurements \cite{Cornell2004}, and surface ion traps \cite{wineland2012}.   A possible solution is to minimize the E-fields by canceling them out.

A convenient surface for applications in HQSs is quartz because of its extensive use in the semiconductor and optics industries.  Despite numerous theoretical and experimental studies of bulk  $\textup{SiO}_2$ \cite{pantelides2013physics, quartzReview2014, louie2005}, the surface properties are not well understood. Recent theoretical work has focused on surface reconstruction and the adsorption of water and graphene \cite{weck2015, Fazzio2011, Okada2011, Wang2006, parker1999}. The $(0001)$ surface has been the subject of recent theoretical interest, partially due to its stability and low surface energy \cite{weck2015, parker1999}.
\begin{figure}[h]
    \includegraphics[width=\columnwidth]{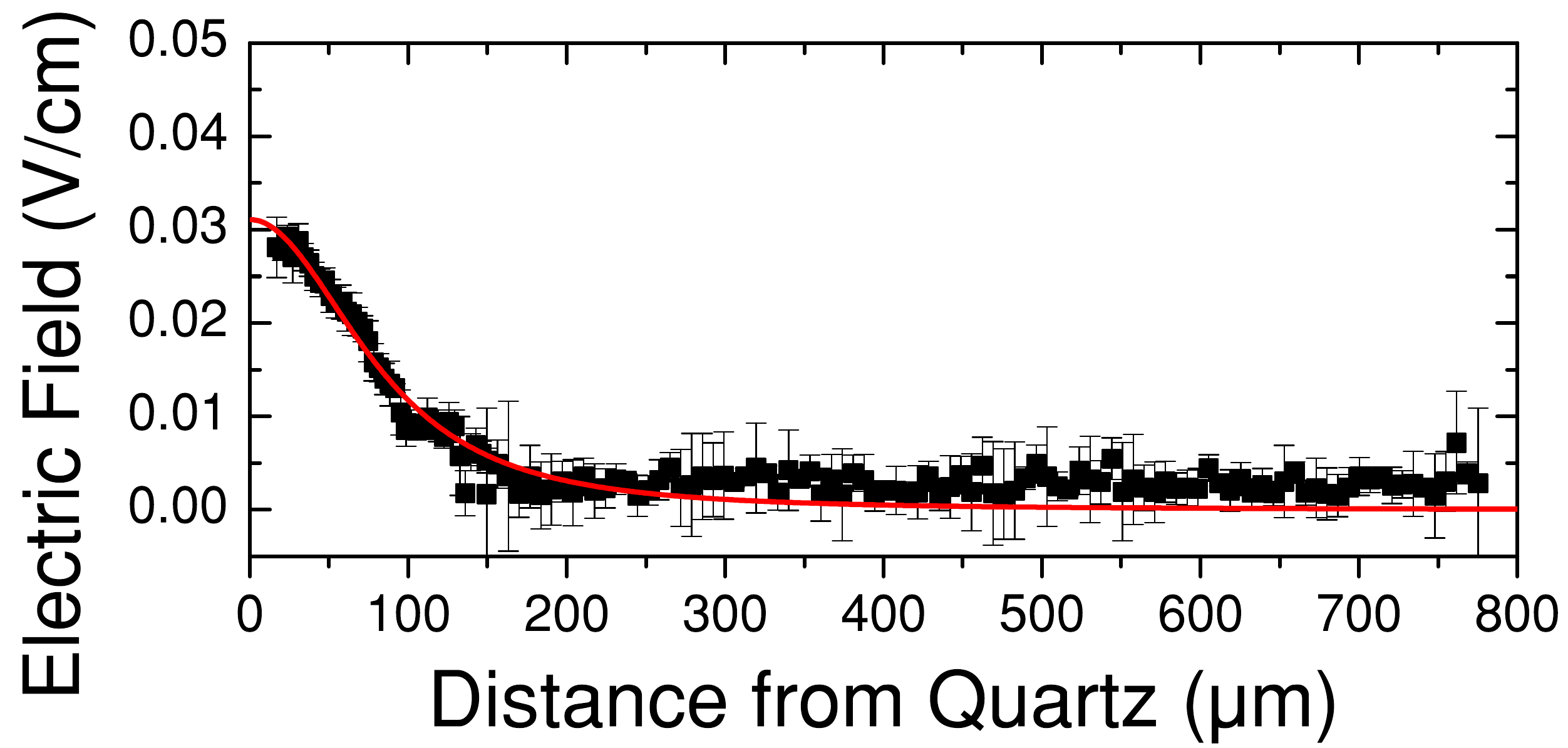}
    \caption{\textbf{Low E-field near the quartz surface.} In the limit of high Rydberg atom (Rb(81$D$)) population the E-field is measured at distances of $\sim 20 - 800 \unit{\mu m}$ from the quartz surface at $T_{\textup{sub}} = 79$ $\celsius$.  The experimental data shown in black is taken in different segments, with changing positions of the coupling beam.  Black points are taken from different pixels on a CCD camera.  The error bars are the standard deviation of the measurement.  The red line is a fit to \Eqnref{adsorbateField}, showing the inhomogeneity of the E-field.  Our calculations indicate that the E-field at $z < 200 \unit{\mu m}$ is caused by the large spacing between electrons on the surface.}
    \label{fig:distgraph}
\end{figure}

In this work, we show that adsorption of Rb atoms on a quartz ($\textup{SiO}_2 (0001)$) surface, contrary to prevailing assumption, can reduce the E-field near the surface, Fig.~\ref{fig:distgraph}. We demonstrate, by appealing to theoretical arguments and {\it ab initio} calculations, that the reduction in E-field is caused primarily by transformation of the quartz into a negative electron affinity (NEA) surface via adsorption of Rb atoms on the surface. A NEA surface can bind electrons, similar to the image potential states on liquid helium (LHe) \cite{coleReview1974, grimes1978review, andreiBook2012}. While the surface repulsion for electrons on LHe is provided by Pauli blocking, the repulsion on quartz occurs because the surface vacuum level dips below the bottom of the conduction band.  We find that the binding of electrons to the surface substantially reduces the E-field above the surface. 

In experiments on atom-adsorbate interactions, using different surfaces, adsorbate E-fields with magnitudes ranging from $ \sim 0.1 - 10 \, \unit{V \,cm^{-1}}$ have been measured at distances of $\sim 10 - 100 \unit{\mu m}$ \cite{Martin2012, hattermann2012detrimental, Spreeuw2010, Haroche2014, Cornell2004}. We measure radically different E-fields depending upon the number of slow electrons produced near the surface.  The E-fields are much smaller when Rydberg atoms near the surface act as a source of slow electrons that can bind to the quartz surface.  We demonstrate that E-fields as small as $30 \unit{mV \, cm^{-1}}$ can be obtained $20 \unit{\mu m}$ from the surface.

A microscopic picture of E-field noise is obtained by considering thermal fluctuations of adsorbate dipole moments near the surface \cite{Safavi2011}.  An adsorbed atom develops a dipole moment as a result of the polarization of the adatom electron cloud in interaction with the surface.  An intuitive model for the dipole created on the surface, $d_0$, is the fractional charge transfer between the adatom and substrate, $\Delta q e$, multiplied by the distance between them \cite{monch2013semiconductor}, $d$, $d_0 = \Delta q e d$.  Calculating $\Delta q$ and $d$ for the Rb-O-terminated quartz system (see Methods), yields $d_0 = 12 \unit{D}$, in good agreement with a more advanced DFT calculation.  $d_0$ is the dipole moment in the limit of low coverage. As the density of adsorbates increases, the E-field from neighboring dipoles reduces the effective dipole moment of each adatom (see Methods).
\begin{figure}[h]
  \includegraphics[width=\columnwidth]{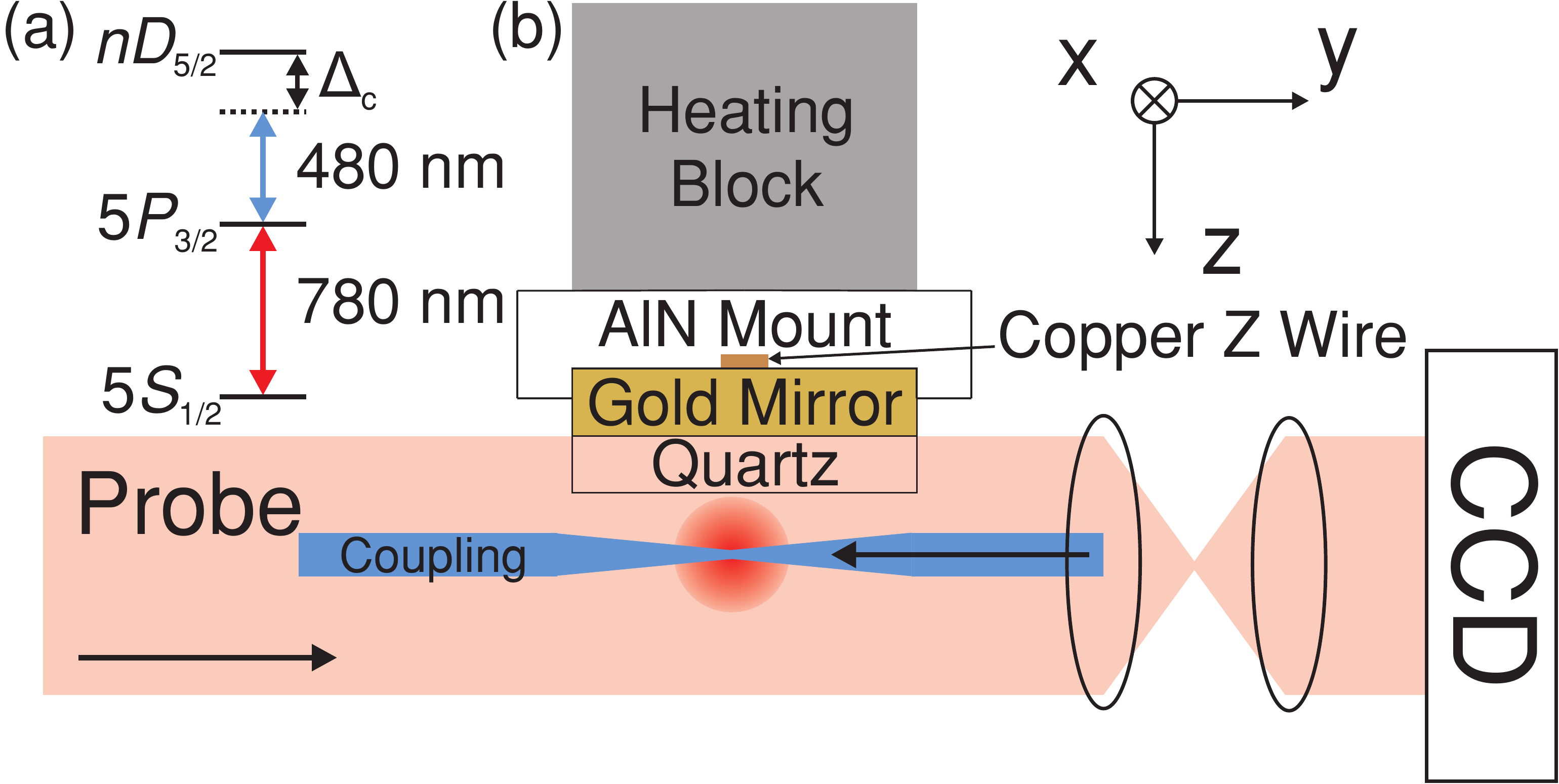}
  \caption{\textbf{Overview of the experiment.} (a) Level scheme for Rydberg EIT used in our experiments.  (b) A schematic of the experimental setup.  Rb atoms are first trapped in a mirror MOT, then transferred into a magnetic trap and transported to the surface.  The probe and coupling beams for Rydberg EIT are overlapped and counterpropagate inside the atom cloud.  The Rydberg EIT signal is observed by analyzing the absorption of the probe beam on a CCD camera.  Heaters are positioned outside of vacuum and control the temperature of the quartz.}
  \label{fig:expSetup}
\end{figure}

Adsorption of a large number of Rb atoms on the quartz surface produces macroscopic E-fields.  At distances far from the surface $z \gg d$, the E-field can be modeled as two finite sheets of charge separated by a small distance \cite{chan2014adsorbate,Spreeuw2010}.  We model the E-field with uniformly charged square sheets of length $L$.  Near the center of the sheets, the E-field is largely perpendicular to the surface,
\begin{equation}
E_z(z) = \frac{2 \sqrt{2} \sigma d(\sigma) L^2}{\pi \epsilon_0 \sqrt{L^2+2z^2}(L^2+4z^2)},
\label{adsorbateField}
\end{equation}
where $\epsilon_0$ is the permittivity of free space, $\sigma$ is the adsorbate density, and $d(\sigma)$ is the coverage dependent dipole moment.  The temperature dependence of $\sigma$ in the limit of low coverage is \cite{dynamicalAdsorption},
\begin{equation}
\dfrac{\sigma/\sigma_0}{1-\sigma/\sigma_0} = C e^{\frac{E_a}{kT_\mathrm{sub}}},
\label{activationEnergy}
\end{equation}
where $\sigma_0$ is the density of adsorbate sites, $E_a$ is the desorption activation energy, $k$ is the Boltzmann constant, and $T_{\textup{sub}}$ is the substrate temperature. Equations (\ref{adsorbateField}) and (\ref{activationEnergy}) relate the adsorbate E-field to $\sigma$ at a given temperature.

\begin{figure}[h]
    \includegraphics[width=\columnwidth]{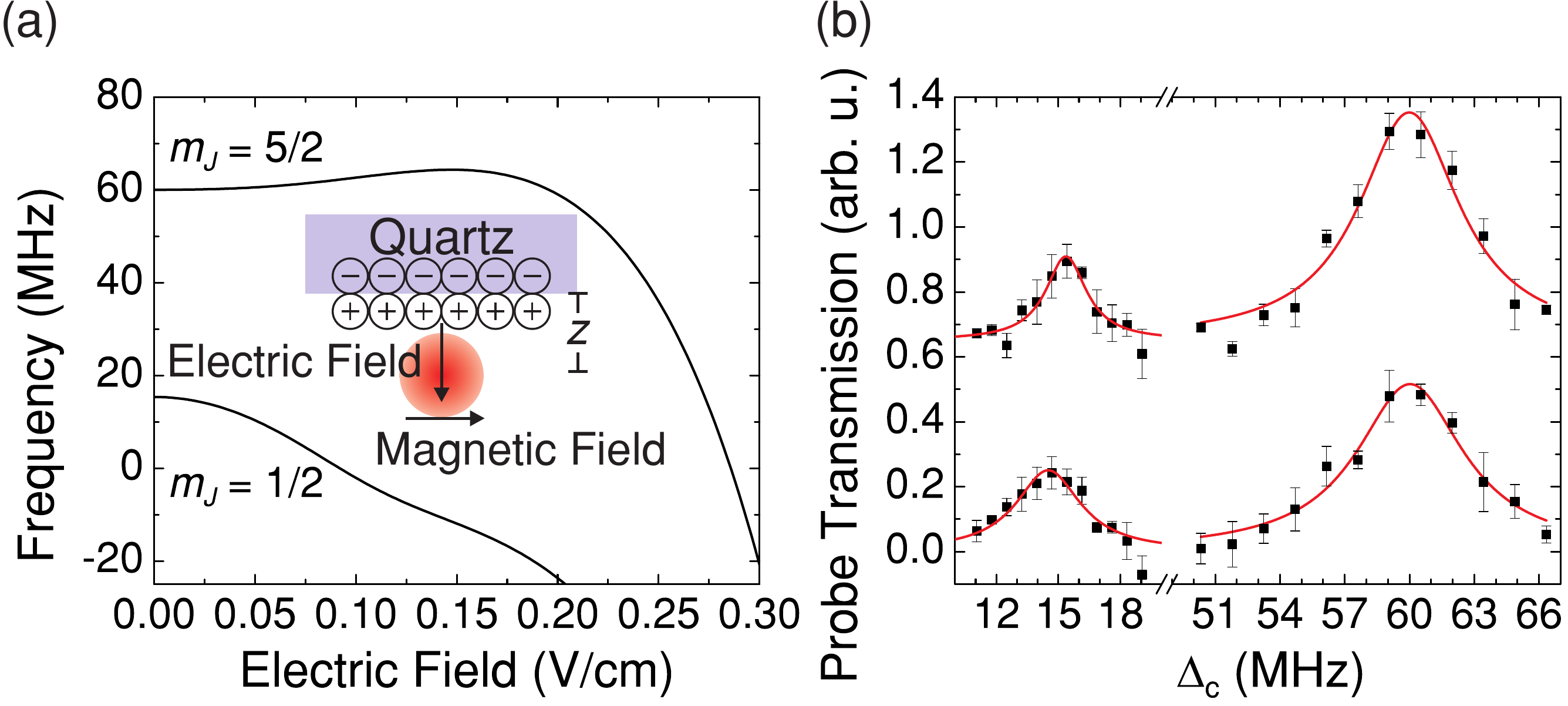}
    \caption{\textbf{Stark shift calculation and Rydberg EIT spectra.} (a) Stark shift for $81D_{5/2}$, $m_J = 5/2$  and $1/2$ states in a 14.3 G magnetic field oriented perpendicular to the E-field.  The inset shows the orientation of the electric and magnetic fields with respect to the quartz surface. (b) EIT spectra taken at 2 different positions $z = 150 \unit{\mu m}$ (upper) and $z = 50 \unit{\mu m}$ (lower) for $81 D_{5/2}$ $m_J=1/2$ (left) and $m_J = 5/2$ (right).  The black points are pixel values of 3 averaged images, and the error bars are the standard deviation of the pixel values.  The red lines are Lorentzian fits to the data.  At $z = 50 \unit{\mu m}$ the $m_J = 1/2$ state is broadened and shifted corresponding to an E-field of $0.02 \unit{V \, cm^{-1}}$.}
    \label{fig:starkshift}
\end{figure}

The E-field is determined by measuring the frequency shift of a Rydberg state, and comparing it to a Stark shift calculation.  The position of the Rydberg state is determined using Rb Rydberg atom electromagnetically induced transparency (Rydberg EIT) \cite{adams2007}.  The energy level scheme for the Rydberg EIT is shown in \Figureref{fig:expSetup}a. When the coupling beam, $480 \unit{nm}$, is on resonance with the upper transition, the atoms are transparent to the 780 nm probe laser.  By analyzing absorption images as a function of coupling laser detuning, the spatial dependence of the E-field is measured with a resolution of $5.5 \unit{\mu m}$.  Stark shifts of two magnetic states for $81D_{5/2}(m_J = 5/2 \mathrm{\, and \,} m_J = 1/2)$ are shown in \Figureref{fig:starkshift}a.  An example of  experimental traces at different $z$ is shown in \Figureref{fig:starkshift}b.

Our experimental setup is shown in \Figureref{fig:expSetup}b.  A mirror magneto-optical trap (MOT) is used to load a Rb magnetic trap $\sim 2 \unit{mm}$ from the quartz surface. After loading the magnetic trap, bias magnetic fields are used to move the atoms close to the surface.  The atoms are released from the magnetic trap and imaged. The atom cloud has approximate dimensions of $1 \times 1 \times 2 \unit{mm}$.

At low Rydberg atom number, the E-field is seen to be homogeneous over the magnetic trap because the EIT signal is not detectably broadened across the extent of the atom sample, $\sim 2 \unit{mm}$.  The variation of the E-field over $z = 200 - 1000 \unit{\mu m}$ is $<0.1 \unit{V}\unit{cm^{-1}}$.  The sensitivity of this measurement is limited by the polarizability of the Rydberg state.  $L$ is estimated to be $10 \unit{mm}$ and is similar in size to other observations \cite{chan2014adsorbate}. 
Disabling the magnetic trap for $\sim 10 \, \unit{minutes}$ did not change the E-field.  Disabling the MOT for the same time period changes the E-field.

\begin{figure}[h]
    \includegraphics[width=\columnwidth]{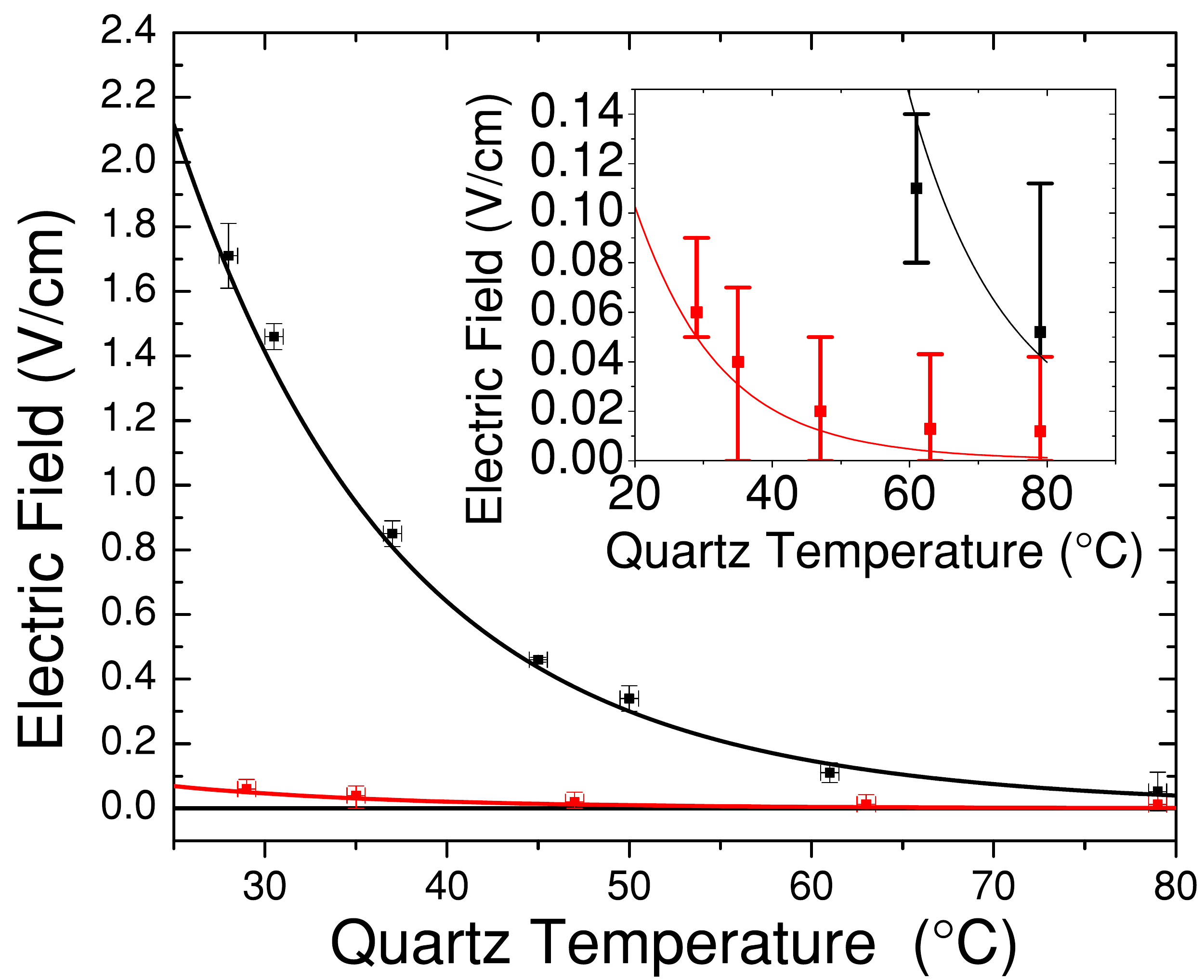}
    \caption{\textbf{Temperature dependence of the E-field.} The measured E-fields due to Rb adsorbates on the (0001) surface of quartz as a function temperature, at a distance of $500 \unit{\mu m}$ from the surface.  The E-fields are calculated by analyzing the frequency shifts of the EIT spectra and comparing them to Stark shift calculations.  The black points are in the limit of low Rydberg atom production.  The black line is a fit to the Langmuir isobar of \Eqnref{activationEnergy}, and yields a desorption activation energy of $E_a = 0.66 \pm 0.02 \unit{eV}$.  The red data points were taken with high Rydberg atom production. The red line is explained in the text.  The horizontal error bars are due to the uncertainty in the temperature of the quartz surface.  The vertical errors bars are the standard deviation of the experimental data.  In the case of high (low) Rydberg atom production the Rabi frequencies of the probe and coupling lasers are $\Omega_p = 2 \pi \times 3.5 (0.5) \unit{MHz}$ and $\Omega_c = 2 \pi \times 4 (4) \unit{MHz}$. }
    \label{fig:tempgraph}
\end{figure}

The adsorbate E-field points away from the surface as confirmed by an external compensating E-field.  The adsorbate E-field is estimated to point perpendicular to the surface, within $15 \degree$, based on the agreement of the overall and differential shifts of different $m_J$ states.  This further justifies the model in \Eqnref{adsorbateField}.

We measured the E-field as a function of the quartz temperature. The results are shown (black) in \Figureref{fig:tempgraph} at $z=500 \unit{\mu m}$.  At $28  \unit{\celsius}$, the E-field is $1.7 \pm 0.1 \unit{V cm^{-1}}$.  Using \Eqnref{adsorbateField}, for a slab of length $L = 10 \unit{mm}$ and $d_0 \sim 12 \unit{D}$, we estimate $\sigma = 4 \times 10^5 \unit{atoms \, \mu m^{-2}}$, yielding an average Rb spacing of $\sim 1.5 \unit{nm}$, and an adatom coverage of $11\% $.  Fitting all values of $\sigma$ to \Eqnref{activationEnergy} with a coverage dependent dipole moment, yields $E_a = 0.66 \pm 0.02\unit{eV}$.  $E_a$ is similar to the $E_a$ for alkali atoms on similar surfaces \cite{Cornell2007, bouchiat1999, stephens1994, Gozzini1992}.

Increasing the Rydberg atom number in either trap dramatically reduces the E-field, by increasing the flux of slow blackbody ionized electrons that can bind to the surface.  The Rydberg atom number can be made larger by increasing the probe laser Rabi frequency.  The temperature dependence of the reduced E-field is shown (red) in  \Figureref{fig:tempgraph}.  For typical parameters shown in \Figureref{fig:tempgraph}, 300 atoms are ionized per experimental sequence.  At $T_{\mathrm{sub}} \sim 28 \unit{\celsius}$, the E-field is reduced by a factor of $\sim 30$.  Reductions in the E-field are observed for all Rydberg states over a range of a principal quantum numbers, $n$. $nS$ and $nD$ states were investigated for $n \sim 40-100$.

The Rydberg atoms are predominately ionized due to blackbody radiation; direct blackbody ionization accounts for $99\%$ of all electrons \cite{beterov2007} at high $n$. For Rb($81D_{5/2}$), the electrons have an average kinetic energy of $10 \unit{meV}$.  Over the range of $n \sim 40 - 100$, the electrons average kinetic energy ranges at $8-15 \unit{meV}$ \cite{Li2004}.

\begin{figure*}[h!t]
\centering
\includegraphics[width= \textwidth]{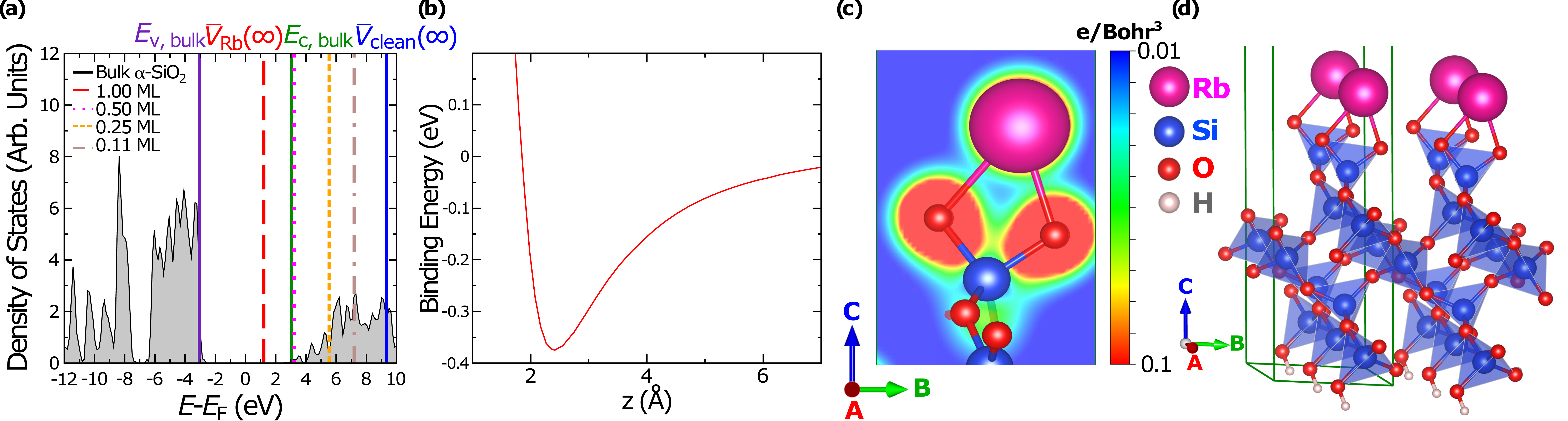}
\caption{\textbf{DFT calculations of Rb on quartz.}  (a) Density of states of bulk $\alpha$-quartz calculated at the GGA/PBE level (see Methods). The Fermi level, $E_F$, top of the valence band, $E_\mathrm{v,bulk}$, bottom of the conduction band $E_\mathrm{c,bulk}$ of bulk $\alpha$-quartz, and the vacuum levels of the $\textup{SiO}_2 \, (0001)$ surface without and with Rb adatom, respectively, $\ \overline{V}_\mathrm{clean}(\infty) \, \mathrm{and} \, \overline{V}_\mathrm{RB}(\infty)$ are labeled. (b) The interaction potential (in eV) for Rb adsorbed on the O-terminated quartz, with z(\AA) normal to the surface. The interaction potential is calculated for a Rb coverage ratio of 1 ML. (c) Charge density map for Rb-SiO$_2$(0001) in the plane of Rb and surface terminated O atoms.  (d) Side view of Rb adsorption on a $\textup{SiO}_2(0001)$ slab.  The Rb atom (pink) is bound to 2 oxygen atoms (red).  The bottom of the slab is passivated by attaching hydrogen atoms (white).  The green axes outline the periodic supercell used in the DFT calculations for 1 ML Rb coverage.}
\label{fig:nea}
\end{figure*}

If the blackbody ionized Rydberg electrons can bind to the surface, they can neutralize the E-field produced by the Rb-adatoms on the surface.  Electrons can bind to a conducting or dielectric surface by binding to their image potential \cite{jackson}. Such bound surface states are ultra-short lived, and rapidly collapse into the bulk.  In LHe, the Pauli repulsion provides the necessary barrier of $\sim 1 \unit{eV}$, preventing the collapse of the electron wave function and leading to the formation of $s$-wave bound states on the surface. In LHe, quantum statistics repel the electrons, allowing for electrons to remain in these bound states for tens of hours at cryogenic temperatures \cite{andreiBook2012}. For adsorption on ordinary surfaces, if the vacuum energy dips below the bottom of the conduction band, a negative electron affinity surface is produced, repelling electrons from the surface.

Amorphous quartz has a positive electron affinity of $0.9 \unit{eV}$ \cite{brorson1985direct}. However, adsorption of atoms can change the substrate surface properties.  The dipole layer created by the adsorbates changes the electric potential at the vacuum-surface interface.  By calculating the electrostatic change in energy of an electron across the surface dipole layer, an estimate of the change in electron affinity, $\Delta \chi$, can be made \cite{Ristein200437} (see Methods). Using $d_0 = 12 \unit{D}$ and $\sigma = 4.2 \times 10^5 \unit{atoms \, \mu m^{-2}}$ at $T = 28 \celsius$, the change in surface electron affinity is  $\Delta \chi = -1.9 \unit{eV}$.  This straightforward approximation suggests that Rb at our densities can shift the vacuum level $\sim 1 \unit{eV}$ below the conduction band, inducing a NEA surface on quartz.  This model shows a NEA up to surface temperatures of $\sim 40 \unit{\celsius}$.

To investigate the adatom-surface on a microscopic level, we performed total-energy calculations for the $(0001)$ surface of quartz with various Rb coverage using spin-polarized density functional theory, as implemented in the Vienna Ab initio Software Package (VASP) \cite{Kresse1996} (see Methods).  The  Rb-quartz interaction potential is shown in \Figureref{fig:nea}b, for a coverage of one monolayer (ML).  On the surface of quartz, the Rb atom is bound to two oxygen atoms.  The concentration of electron density around the two oxygen atoms is shown in \Figureref{fig:nea}c. A side view showing Rb adsorption is shown in \Figureref{fig:nea}d.  In the potential, the lowest bound state has an energy of $E_b=0.35 \unit{eV}$.  For the lower coverages investigated experimentally, our DFT calculations show an increase of $E_b$ by $\sim 1.4$.  The calculated $E_b$ is comparable in magnitude with the measured $E_a$, and is consistent with the expectation $E_b \leq E_a$ \cite{brillson2010surfaces}.

We calculated the electronic density of states (DOS) for the bulk $\alpha$-quartz and the shift of the vacuum energy with varying amounts of Rb coverage using DFT.  The results are shown in \Figureref{fig:nea}a and the details of the calculations are in the Methods section. The Fermi level, $E_F$ is set equal to zero, and lies in the middle of the band gap, between the top of the valance band, $E_\mathrm{v,bulk} = -3.05 \unit{eV}$, and the bottom of the conduction band, $E_\mathrm{c,bulk} = 3.05 \unit{eV}$. As shown in \Figureref{fig:nea}a, the vacuum level for the clean surface, $\overline{V}_\mathrm{clean}(\infty)$, has a positive electron affinity, consistent with experiment \cite{brorson1985direct}.  However, adsorbing Rb on the surface shifts the vacuum level downward.  NEA is induced around $0.5 \unit{ML}$.  The DFT and the straightforward electrostatic calculations, both show that the vacuum level shifts by several electron volts with {\it only} a modest amount of Rb coverage.  The remaining discrepancy may be resolved with further improvements in DFT \cite{Yang2008,cohen2011challenges}.  More knowledge of the experimental surface including the Rb adsorbate structure will also help to guide the calculations.

We model the electrons as a uniformly charged square sheet of length $L$, that overlays the adsorbate layers with $L = 10 \unit{mm}$ at $z = 0$.  The resulting E-field is a sum of the E-fields from the adsorbates and electrons, $E_\textup{tot} = E_\textup{ads}+E_\textup{ele}$.  After requiring the $E_{\mathrm{tot}} = 0$ at $z=0$,  $E_\textup{tot}(z=500 \unit{\mu m})$ is plotted in \Figureref{fig:tempgraph} (red).  The near exact fit to data is an indication that the reduction in the E-field is due to the formation of a NEA surface for Rb-SiO$_2$.

For high temperatures and high Rydberg population the E-field produced by the surface is low.  The measured E-field as a function of $z$ at $T = 79 \unit{\celsius}$ is shown in \Figureref{fig:distgraph}.  For $z > 200 \unit{\mu m}$ the E-field is negligible within the error.  At $z < 200 \unit{\mu m}$ the E-field increases to $\sim 30 \unit{mV \, cm^{-1}}$.  Under these conditions, we estimate a surface electron density of $\sim \, 10 \, \unit{electrons \, mm^{-2}}$.  For $z < 200\,\mu$m, approximating the electrons as a sheet with uniform charge breaks down since the spacing between electrons is larger than $z$.  The spectral width of the EIT resonance for 81 $D_{5/2}$, $(m_J = 1/2)$ increases from $2 \unit{MHz}$ far from the surface to $\sim 4 \unit{MHz}$ at $z \leq 50 \unit{\mu m}$.  We attribute this broadening to the inhomogeneity of $E_\textup{ele}$ near the surface.  The data in \Figureref{fig:distgraph} is fit to \Eqnref{adsorbateField}, and shows that the residual E-field can be modeled as a dipole patch of adsorbates, with $L \sim 200 \unit{\mu m}$. $L$ is approximately equal to the estimated electron spacing $\sim 300 \unit{\mu m}$.

\begin{figure}[h]
    \includegraphics[width=\columnwidth]{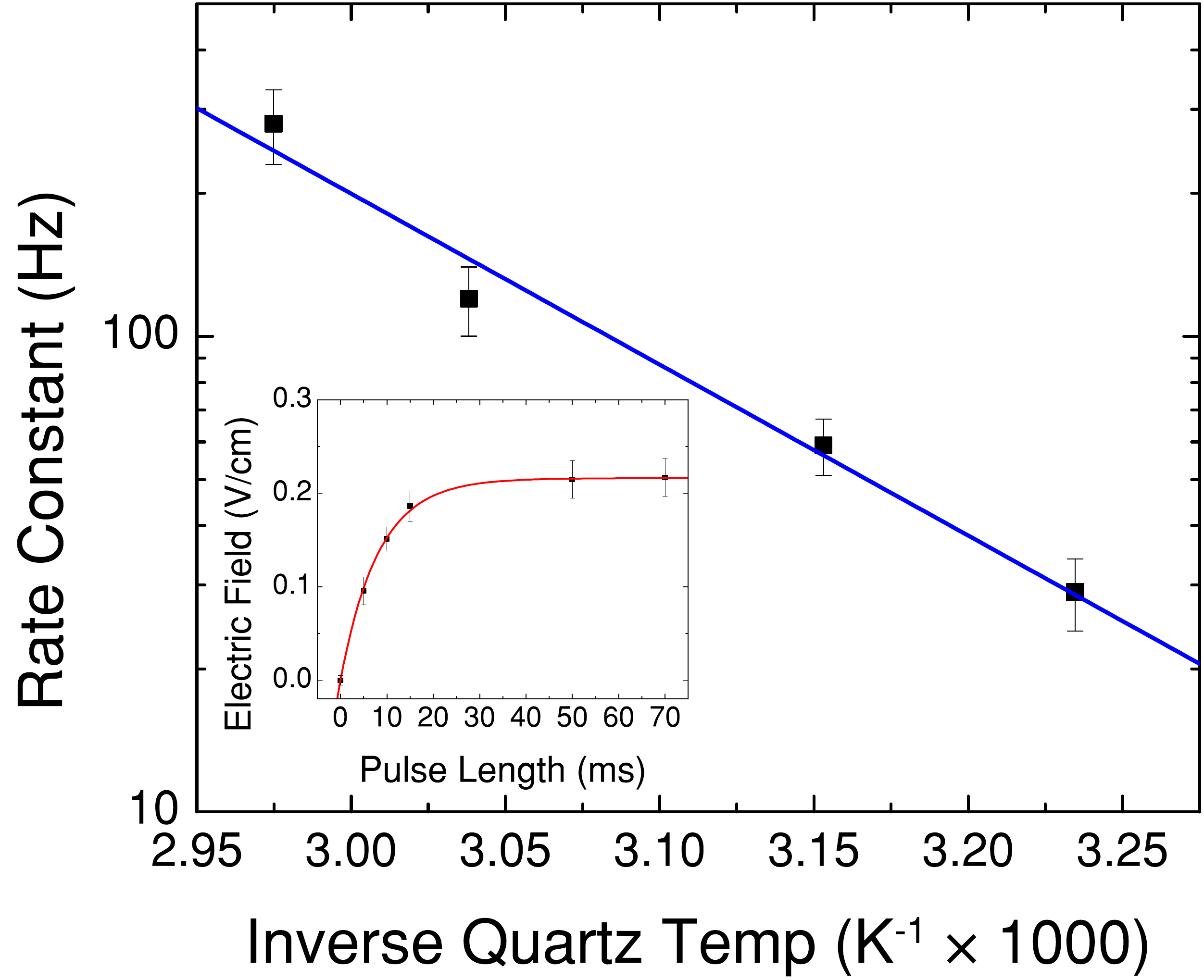}
    \caption{\textbf{Photodesorption of electrons with UV light.}  A short pulse of 400 nm light is incident upon the quartz surface saturated with electrons.  The inset shows the increase of the E-field with the pulse duration at $T_\textup{sub} = 56 \unit{\celsius}$.  The black points are experimental data, and the error bars are the standard deviation of the measurement.  The red line is a fit to $E(t) = A(1-e^{-bt})$, where the photodesorption rate constant $b$ is extracted from the fit.  The main figure shows $b$ measured for different quartz temperatures.  The error bars are the uncertainty in the fit of the data to $b$.  The blue line is a fit of the data to the Arrhenius equation, with an activation energy of $0.7 \pm 0.07 \unit{eV}$.}
    \label{fig:uv}
\end{figure}

We can remove electrons from the surface using 400 nm light generated with a light emitting diode (LED) array.  We start with the surface saturated with electrons.  The LEDs are pulsed on for a variable time while atoms are being loaded into the MOT.  The intensity of the pulse is kept small to avoid light induced desorption of Rb. The MOT fluorescence is monitored to verify this condition.  The E-field is measured using Rydberg atom EIT.  An example measurement for $T = 56 \unit{\celsius}$, is shown in the inset of \Figureref{fig:uv}.  Assuming a direct process, the data is fit to an exponential function, $E(t) = A(1-e^{-bt})$.  $A$ is the largest value of the E-field measured and $b$ is the photodesorption rate constant.  $b$ is measured over a range of temperatures, shown in the main part of \Figureref{fig:uv}.  The photodesorption rate constant has an Arrhenius behavior, with an activation energy of $0.7 \pm 0.07 \unit{eV}$.  The activation energy is similar to the $E_a$ suggesting the removal mechanism of Rb on quartz is dependent on Rb coverage.  The Rb coverage is the variable that affects the energy levels most strongly.  It is unknown if the electrons are detached from or tunnel into the surface.  The photodetachment mechanism is the subject of further investigation.

Over the temperature range investigated, $28 \celsius < T< 80 \celsius$, the Rb-quartz system can bind electrons for several hours.  The small E-fields have been repeatably measured many times for over a year, yielding the same results within experimental error.  The thermal wavelength of an electron at $28 \celsius$ is $4.3 \unit{nm}$, indicating that the electron is not localized to one Rb adsorbate. We believe that the single crystal nature of the quartz and small surface roughness, $<5 \unit{\AA}$, plays an important role in the uniformity of the Rb adsorbates and electron binding.  There are few steps and structures on the surface to nucleate Rb clusters, leading to the homogeneous fields we observe.  We have done some simulations investigating whether the dipole potential from a patch of adsorbates or the image potential is responsible for binding the electrons to the surface.  Our results show that binding is due to the image potential of the electron, and the patch potential slightly shifts the image potential.

In summary, we have measured the activation energy of Rb on the $(0001)$ surface of quartz and shown the onset of a NEA surface capable of binding electrons upon Rb adsorption.  Reducing E-fields on a quartz surface by making quartz a NEA surface by Rb adsorption is a promising pathway for coupling Rydberg atoms to surfaces.  Further work can be directed towards measurements of other surface orientations and dielectrics, as well as investigating the behavior at cryogenic temperatures.  The properties of the electrons, including binding energy, mobility, and effective mass, are the subject of future work.

\section{Acknowledgments}
Sandia National Laboratories is a multi-program laboratory managed and operated by Sandia Corporation, a wholly owned subsidiary of Lockheed Martin Corporation, for the U.S. Department of Energy's National Nuclear Security Administration under contract DE-AC04-94AL85000. This work was supported by the DARPA Quasar program by a grant through ARO (60181-PH-DRP), AFOSR (FA9550-12-1-0282), NSF (PHY-1104424) and an NSF grant through ITAMP at the Harvard-Smithsonian Center for Astrophysics. The authors thank Tilman Pfau for useful discussions.

\section{Author contributions}
J. A. S. and J. P. S. performed experiments and analyzed the data.  E. K. and P. F. W. performed the DFT calculations.  S. T. R. and H. R. S. performed surface binding calculations.  All authors contributed to the preparation of the manuscript.
\section{Competing financial interests}
The authors declare no competing financial interests.

\bibliographystyle{bibsty}

\pagebreak


\setcounter{equation}{0}
\setcounter{figure}{0}
\setcounter{table}{0}
\setcounter{page}{1}
\makeatletter
\renewcommand{\theequation}{\arabic{equation}}
\renewcommand{\thefigure}{\arabic{figure}}
\renewcommand{\bibnumfmt}[1]{[#1]}
\renewcommand{\citenumfont}[1]{#1}
\section{Methods}
\subsection{Experimental Details}
About $\sim$ $2 \times 10^6$ $^{87}$Rb atoms are captured in a mirror MOT $\sim 2 \unit{mm}$  from the surface. The quadrupole magnetic field for the MOT is generated by external coils. The MOT coils are shut off and the atoms are optically pumped into the $F=2$, $m_F = 2$ state and loaded into a Ioffe-Pritchard magnetic trap. About 2/3 of the atoms are transferred into the magnetic trap.  The magnetic fields for the Ioffe-Pritchard trap are produced by a millimeter sized Z-wire situated $1.2 \unit{mm}$ above the quartz surface.  Bias fields for the trap are created by external coils.  After loading the magnetic trap, the bias fields are used to move the atoms close to the quartz surface.  The atoms are released from the magnetic trap and imaged.  After release, the atoms have a peak atomic density of $3 \times 10^9 \unit{cm^{-3}}$ and a temperature of $100 \unit{\mu K}$.  The atom cloud has approximate dimensions of $1 \times 1 \times 2 \unit{mm}$.  The background pressure in the chamber is $3 \times 10^{-10} \unit{Torr}$.
We use a $20 \times 20 \times 0.5 \unit{mm}$ piece of single crystal z-cut $(0001)$ quartz with surface roughness of $<5 \unit{\AA}$. The quartz is mounted to a gold mirror of thickness $\sim 700 \unit{\mu m}$ that is used for the mirror MOT.  Heaters outside the chamber heat the whole assembly.  The temperature of the quartz is monitored with a thermocouple, located far away ($\sim 14 \unit{mm}$) from the magnetic trap.  The gold mirror for the mirror MOT is epoxied to an aluminum nitride mount that also holds the copper Z-wire.  The aluminum nitride mount is in thermal contact with an aluminum block. The quartz is mounted on the gold mirror. The entire assembly is mounted on copper feedthroughs so that it can be heated outside the vacuum chamber.

The 780 nm light is produced by an extended cavity diode laser (ECDL) and the 480 nm light is generated by a home-built frequency doubling cavity of an amplified 960 nm ECDL.  The 780 nm and 960 nm lasers are stabilized to a ultrastable Fabry-Perot cavity (Stable Laser Systems). The transition frequencies in the experimental chamber are referenced to a vapor cell located on a different optical table.  The Stark shifts and the magnetic field were verified by comparing the observed shifts in the vapor cell with an applied magnetic field. The probe and coupling beams for the EIT detection are counter-propagating with pulse lengths of $150 \unit{\mu s}$.  The probe beam is collimated with a waist of $4 \unit{mm}$.  The coupling beam is focused through the atom cloud with a waist of $50 \unit{\mu m}$.  The probe beam is $\sigma^+$ polarized and the coupling beam is linearly polarized in the x-direction.

\subsection{Summary of DFT Calculation}
The exchange correlation energy was calculated using the generalized gradient approximation with the parameterization of Perdew, Burke, and Ernzerhof \cite{Ernzerhof1996} (PBE).  The interaction between valence electrons and ionic cores was described by the projector augmented wave (PAW) method \cite{blochl1994, Kresse1999}.  The Rb $(4s^2,4p^6,5s^1)$, Si $(3s^2,3p^2)$, and O $(2s^2,2p^4)$ electrons were treated explicitly as valence electrons in the Kohn-Sham (KS) equations and the remaining cores were represented by PAW pseudopotentials.  The KS equations were solved using the blocked Davidson iterative matrix diagonalization scheme \cite{WilsonBook} followed by the residual vector minimization method.  The plane-wave cutoff energy for the electronic wave functions was set to $500 \unit{eV}$, ensuring the total energy of the system to be converged to within $1 \unit{meV/atom}$.

Structural optimization was carried out with periodic boundary conditions applied using the conjugate gradient method, accelerated using the Methfessel-Paxton Fermi level smearing \cite{Paxton1989} with a width of $0.1 \unit{eV}$.  The total energy of the system and Hellmann-Feynman forces acting on atoms were calculated with convergence tolerances set to $10^{-3} \unit{eV}$ and $0.01 \unit{eV/ \AA}$, respectively.  Structural optimizations and properties calculations were carried out using the Monkhorst-Pack special \emph{k}-point scheme \cite{Monkhorst1976} with $5\times5\times5$ and $5\times5\times1$ meshes for integrations in the Brillouin zone (BZ) of $\textup{SiO}_2$ bulk and slab systems, respectively.

The simulation supercell consisted of periodic $(2\times2)$ $\textup{SiO}_2 \, (0001)$ slab surfaces separated by $30.00 \unit{\AA}$ vacuum layers. O-terminated $\textup{SiO}_2 \, (0001)$ surfaces were built by cleaving the relaxed bulk structure of $\alpha$-quartz and consisted of 5 $\textup{SiO}_2$ bilayers, with the back side of the slabs passivated by addition of hydrogen atoms.  In the structural optimization calculations, only the top two $\textup{SiO}_2$ bilayers were allowed to relax. Although a large vacuum region was used between periodic slabs, the creation of dipoles upon adsorption of atoms on only one side of the slab can lead to spurious interactions between the dipoles of successive slabs.  In order to circumvent this problem, a dipole correction was applied by means of an artificial dipole layer placed in the vacuum region following the scheme introduced by Neugebauer and Scheffler \cite{Scheffler1992}.  A similar computational approach was used previously to investigate dipole formation upon Cs adsorption onto a graphene-veiled $\textup{SiO}_2 \, (0001)$ slab surface \cite{weck2015_S}.

$\overline{V}(\infty)$, the plane-averaged electrostatic potential in the vacuum at a distance where the microscopic potential has reached its asymptotic value, was obtained from a self-consistent electronic structure calculation using a plane wave basis set of the electrostatic potential $V(x,y,z)$ on a grid in real space.  Assuming that the surface normal is oriented along the z-axis, one can define a plane averaged potential:
\begin{equation}
\overline{V}(z) = \frac{1}{A}\int\int_\mathrm{cell} V (x,y,z)dxdy,
\end{equation}
where $A$ is the supercell surface area.  The asymptotic value $\overline{V}(\infty)$ can be extracted by plotting the variation of $\overline{V}$ as a function of $z$.  It should be noted that the major contribution to the surface dipole results from the charge reordering associated with the formation of the chemical bonds between the surface and the adatoms.  This contribution is foremost determined by the nature of the chemical bonds, but can also be modified by the packing density of the adatoms.

In order to quantify the dipole moment created by adsorption of a Rb adatom on $\textup{SiO}_2 \, (0001)$, partial charges were calculated with Bader charge partitioning \cite{bader1990atoms}.  At a coverage, $\sigma/\sigma_0 = 0.11$, the partial charge transferred from one Rb adatom to the surface is $0.947$.  Since the calculated equilibrium $d$(Rb-O) bond distance is $2.79 \unit{\AA}$, the resulting local dipole moment is $12.7 \unit{D}$.

\subsection{Electrostatic Calculations}
We estimated the dipole moment of a Rb atom adsorbed on quartz as using $d_0 = \Delta q d$, as described in the main text. $\Delta q$ is the amount of ionic character of the bond, and is calculated in terms of Pauling electronegativities $X_\textup{A}$ and $X_\textup{B}$ for the Rb and the surface respectively \cite{monch2013semiconductor_S},
\begin{equation}
\Delta q = 0.16|X_A-X_B|+0.035|X_A-X_B|^2.
\label{eq:deltaq}
\end{equation}
The DFT calculations show that the Rb is bound to two oxygens, and the total dipole moment is calculated as the vertical contribution from each Rb-O bond.  The effective electronegativity of each oxygen, $X_\textup{B}$, is calculated by using the geometric mean of the electronegativities of 1 Si atom and 2 O atoms, $X_\textup{B} = (1.90\times3.44\times3.44)^{1/3} = 2.8$.  This results in $\Delta q = 0.46$.  $d$ can be estimated by adding the covalent radii of the atoms,
\begin{equation}
d = r_\textup{cov}^\textup{ads} + r_\textup{cov}^\textup{sub} = 2.79 \unit{\AA}.
\label{eq:rad}
\end{equation}
The resulting dipole moment for each Rb-O bond is, $d_1 = 6.2 \unit{D}$.  The $z$ component of the vertical dipole $d_{1z} = 5.8 \unit{D}$, is calculated using the geometry of the system with the O-O separation of $2.363 \unit{\AA}$. The total dipole moment in the $z$ direction is $d_0=12 \unit{D}$.  The numbers used for the covalent radii and electronegativities are found in \cite{CRC2015}.  This simple calculation supports the DFT result.

The change in the electron affinity can be estimated from \cite{Ristein200437_S},
\begin{equation}
\Delta \chi = -\frac{e d(\sigma)\sigma}{\epsilon_0}
\end{equation}
where $\sigma$ is the density of adsorbates on the surface and $d(\sigma)$ is the density dependent dipole moment,
\begin{equation}
d(\omega) = \frac{d_0}{(1 + 9 \alpha_\textup{ad} \sigma^{3/2})}.
\end{equation}
$\alpha_\textup{ad}$ is the polarizability of the adatoms.  Using the calculated dipole moment, $d_0 = 12 \unit{D}$, and experimental values for $\sigma = 4.2 \times 10^5 \unit{atoms \, \mu m^{-2}}$ at $T = 28 \celsius$, $\Delta \chi = -1.9 \unit{eV}$. Through adsorption of cesium, materials such as diamond \cite{polyakov1998effects, Jackman1997} and gallium nitride \cite{eyckeler1998negative} can shift their surface electron affinity from positive to negative in a similar fashion.


\end{document}